\documentclass{JAC2003}

\usepackage{graphicx}
\providecommand{\PNNRESULT}{\mbox{  $(1.57^{+1.75}_{-0.82})\times 10^{-10}$}}
\def\Br{{\cal B}}

\newcommand{\KL}{\mbox {$ K^0_{\rm L}$ }}          
\providecommand{\PZ}{\mbox{$\pi^0$}}
\newcommand{\KEPG}{\mbox {$ \pi^\pm e^\mp\nu\gamma$ }}           
\newcommand{\KPPP}{\mbox {$ \pi^+\pi^-\pi^0$ }}          
\newcommand{\KPGG}{\mbox {$ \pi^0\gamma\gamma$ }}          
\newcommand{\KGG}{\mbox {$ \gamma\gamma$ }}          
\newcommand{\KZZZ}{\mbox {$ \pi^0\pi^0\pi^0$ }}          
\newcommand{\KEPP}{\mbox {$ \pi^0\pi^\pm e^\mp\nu$ }}        

\newcommand{\BR}{\mbox{${\cal B}$}}          
\newcommand{\ESTAR}{\mbox{$E_\pi^*$}}          
\newcommand{\CHISQ}{\mbox{$\chi^2$}}          
\newcommand{\EMISS}{\mbox{$E_{\rm MISS}$}}          
\newcommand{\DELTAE}{\mbox{$|E_{1\gamma}^*-E^*_{2\gamma}|$}}          
\newcommand{\MGG}{\mbox{$M_{\gamma\gamma}$}}          

\begin{document}
\title{\begin{boldmath} $ {\rm K}\to\pi\nu\bar\nu$ \end{boldmath} at Hadron Machines}

\author{David E. Jaffe, BNL, Upton, NY, USA}

\maketitle

\begin{abstract}
 The results and goals of experiments E787, E949, CKM and KOPIO 
on the measurement of the branching fractions of 
$ {\rm K}^+\to\pi^+\nu\bar\nu$	and
$ {\rm K}^0_{\rm L}\to\pi^0\nu\bar\nu$ 
are presented. 
\end{abstract}

\section{Introduction}

 The branching fractions of $ {\rm K}^+\to\pi^+\nu\bar\nu$	and
$ {\rm K}^0_{\rm L}\to\pi^0\nu\bar\nu$ belong to a small set of measurable
quantities that have a precise and unambiguous relation to the
fundamental parameters of the standard model (SM). Due to the
large top mass, the decays $ {\rm K}\to\pi\nu\bar\nu$  are sensitive to 
 the product $V_{\rm td}V^*_{\rm ts}$ of the CKM elements
that quantify $t\to d$ and $t \to s$ transitions. In terms of the
unitarity triangle (UT) that represents the CKM matrix in the complex plane,
the height is proportional to $ \sqrt{\Br({\rm K}^0_{\rm L}\to\pi^0\nu\bar\nu)}$ 
and the length of one side is proportional to
 $ \sqrt{\Br({\rm K}^+\to\pi^+\nu\bar\nu)}$. The observation of these decays and their branching
fractions is the goal of four experiments at hadronic machines:
E787, E949, CKM ( $ {\rm K}^+\to\pi^+\nu\bar\nu$) and KOPIO ($ {\rm K}^0_{\rm L}\to\pi^0\nu\bar\nu$).
The KEK experiment E391a will also study the
$ {\rm K}^0_{\rm L}\to\pi^0\nu\bar\nu$ and is discussed in these proceedings.

\section{E787}

Experiment E787 at the Alternating Gradient Synchrotron (AGS) of 
Brookhaven National Laboratory (BNL) finished data collection in 1999 and 
has observed 2 candidates for $ {\rm K}^+\to\pi^+\nu\bar\nu$ upon an
estimated background of $0.15\pm0.05$. The probability that the two candidates are 
due to background is 0.02\%. These observations imply
$\Br({\rm K}^+\to\pi^+\nu\bar\nu) = \PNNRESULT$~\cite{ref:e787.pnn1} which
is statistically consistent with the SM prediction
of  $(0.7\pm0.2)\times 10^{-10}$~\cite{ref:sm.pred}
albeit with a central value tantalizingly twice the expectation.

\begin{figure}[htb]
\centering
\includegraphics*[width=65mm]{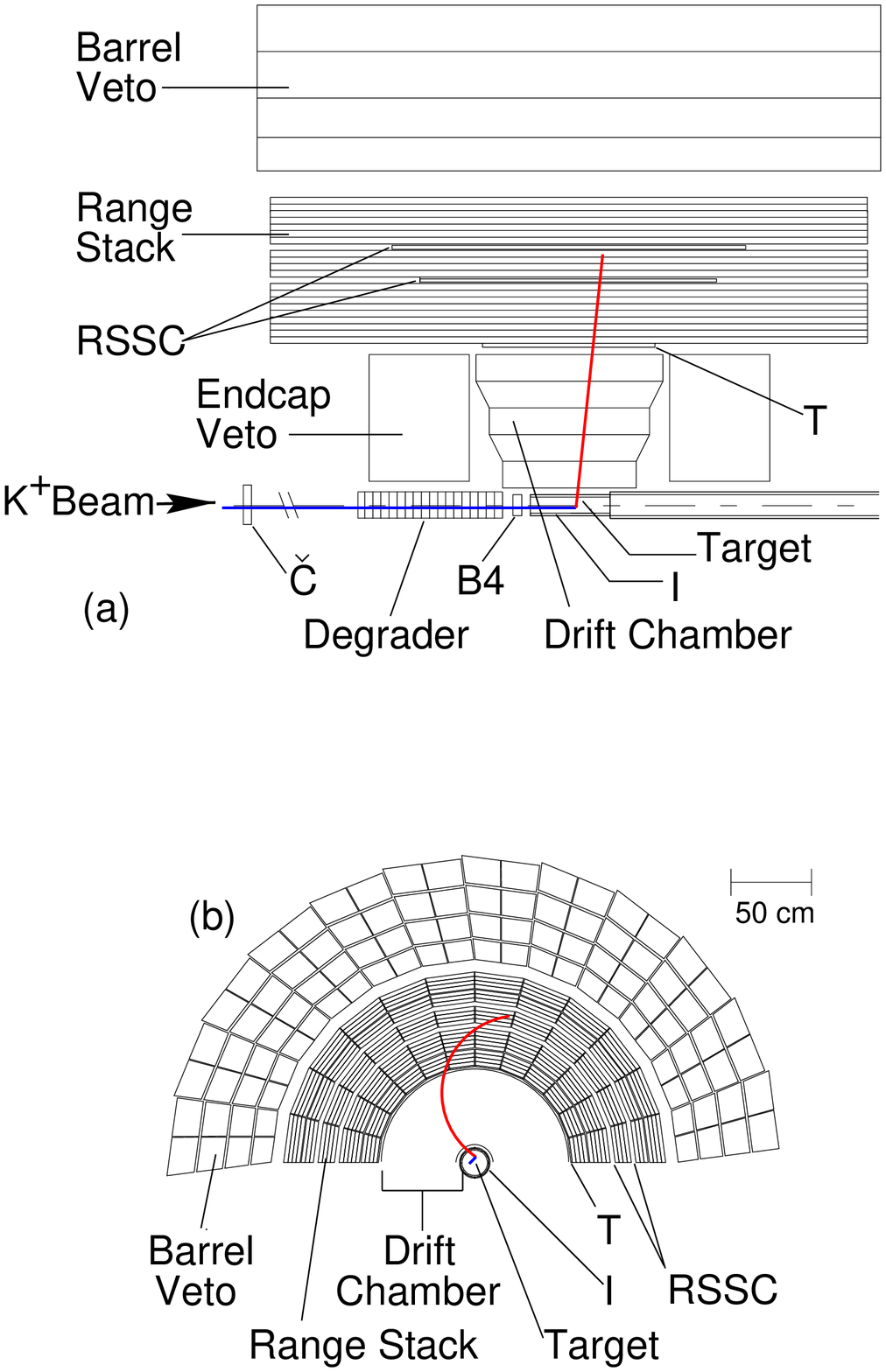}
\caption{Schematics of the E787 detector: (a) plan view, (b) beam's eye view.
Only the upper half of the detector is shown.}
\label{fig:e787.det}
\end{figure}

 A schematic of the E787 detector is shown in Figure~\ref{fig:e787.det}. A $\sim700\ {\rm MeV}/c$ 
beam with ${\rm K}^+/\pi^+ \approx 4$ passes through threshold Cherenkov counters, wire
chambers, a degrader and a plane of hodoscopes before stopping in a scintillating
fiber target. Outgoing $\pi^+$ from ${\rm K}^+$ decays in the range $45^\circ$ to $135^\circ$ with 
respect to the initial  ${\rm K}^+$ direction traverse a low mass drift chamber and come to rest in 
the range stack (RS) of plastic scintillator bars. The RS is surrounded by a non-projective 
lead-scintillator barrel veto approximately 13 radiation lengths ($X_0$) thick.
Pure CsI detectors $\sim 14 X_0$ thick perform a similar function in the end caps. Additional 
lead-scintillator veto counters are inserted in the beam region to improve 
hermeticity.

 To observe the  $ {\rm K}^+\to\pi^+\nu\bar\nu$ decay at the SM predicted rate
of  $(0.7\pm0.2)\times 10^{-10}$~\cite{ref:sm.pred}, E787 needed to suppress
backgrounds by a factor of $10^{11}$. This goal was accomplished by independent measurements of
the momentum (P), range (R) in plastic scintillator and energy (E)
of the  $\pi^+$ . The incoming   ${\rm K}^+$  is positively identified
by Cherenkov light, $dE/dx$ and range in the target. The entire $\pi^+\to\mu^+\to e^+$
decay chain is detected in the RS 
for positive $\pi^+$ identification and is augmented by $dE/dx$ measurements. All active
elements of the detector are used to veto on extra neutral or charged particles.

 The E787 analysis strategy is summarized below:

\begin{itemize}

\item A priori identification of background sources.

\item Suppress each background source with at least two
independent cuts.

\item Backgrounds cannot be reliably simulated: measure with
data by inverting cuts and measuring rejection taking
any (small) correlations into account.

\item To avoid bias, set cuts using 1/3 of data, then
measure backgrounds with remaining 2/3 sample.

\item Verify background estimates by loosening cuts and comparing
observed and predicted rates.

\item Use MC to measure geometrical acceptance for  $ {\rm K}^+\to\pi^+\nu\bar\nu$. Verify by
measuring $\Br({\rm K}^+\to\pi^+\pi^0)$.

\item ``Blind'' analysis. Don't examine signal region until 
all backgrounds are verified.

\end{itemize}

\begin{figure}[htb]
\centering
\includegraphics*[width=65mm]{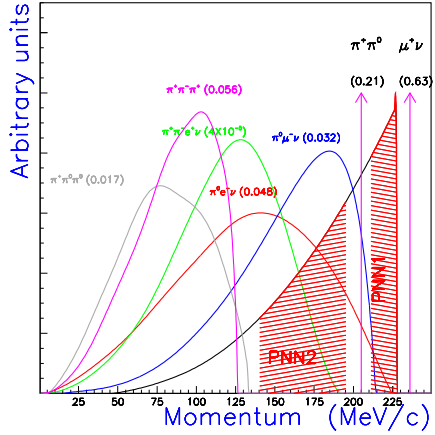}
\caption{Momentum spectra of the $\pi^+$ from the main  ${\rm K}^+$ decay modes.
The range of the search regions above (pnn1) and below (pnn2) the
$K_{\pi2}$ peak are indicated.}
\label{fig:top7}
\end{figure}

\begin{figure}[htb]
\centering
\includegraphics*[width=65mm]{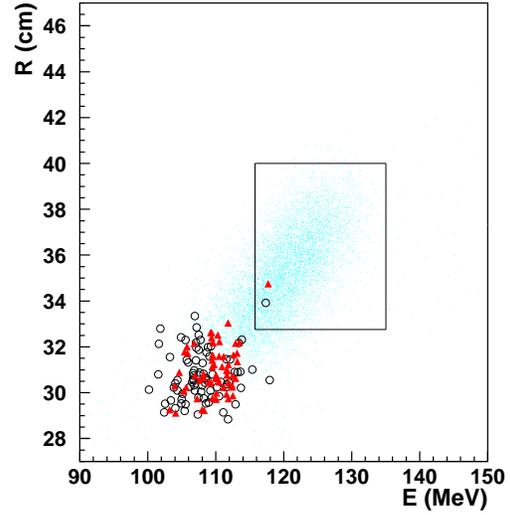}
\caption{The range(cm) vs energy(MeV) of  $ {\rm K}^+\to\pi^+\nu\bar\nu$ 
candidates after all other cuts are applied. The box represents the signal
region. The light dots represent the signal distribution from simulation. The
triangles and circles represent the data from the 1995-7 and 1998 running periods, 
respectively.}
\label{fig:pnn1}
\end{figure}

 E787 searched for  $ {\rm K}^+\to\pi^+\nu\bar\nu$ in two distinct
kinematic regions --- above and below the ${\rm K}^+\to\pi^+\pi^0$ ($K_{\pi2}$)
 peak (Figure~\ref{fig:top7}). The backgrounds in the higher momentum
region, dubbed pnn1, are due to the two-body decays $K_{\pi2}$
and  ${\rm K}^+\to\mu^+\nu$ ($K_{\mu2}$), beam particles scattering into the
RS and the charge exchange (CEX) reaction  $K^+n\to K^0p$, $K^0_L\to\pi^+\ell^-\nu$.
The CEX background is estimated from a combination of data and simulation; 
the other backgrounds are estimated from data as outlined above.
The E787 results in the pnn1 region are given in Table~\ref{tab:pnn1}
and Figure~\ref{fig:pnn1} and result in 
$\Br({\rm K}^+\to\pi^+\nu\bar\nu) = \PNNRESULT$~\cite{ref:e787.pnn1}.

\begin{table}[hbt]
\begin{center}
\caption{E787 results for the pnn1 search region for the 1995-7 and 1998
running periods.
N(K) is the number of stopped $ {\rm K}^+$,
Acc. is the acceptance, 
Sens. is the single event sensitivity, 
Cand. is the number of observed signal candidates
and $\Br$ is $\Br({\rm K}^+\to\pi^+\nu\bar\nu)$.
}
\begin{tabular}{|l|c|c|}
\hline
Bkgd		& 1995-97	& 1998	\\
\hline
$K_{\pi2}$	& $0.03\pm0.01$	& $0.012{}^{+0.003}_{-0.004}$\\
$K_{\mu2}$	& $0.02\pm0.01$ & $0.034{}^{+0.043}_{-0.024}$\\
Beam		& $0.02\pm0.02$ & $0.004\pm 0.001$	\\
CEX		& $0.01\pm0.01$ & $0.016{}^{+0.005}_{-0.004}$\\
\hline
Total 		& $0.08\pm0.03$	& $0.066{}^{+0.044}_{-0.025}$\\
\hline
\hline
N(K)		& $3.2\times 10^{12}$ 	& $2.7\times 10^{12}$	\\
Acc.		& $0.0021(1)(2)$ 	& $0.00196(5)(10)$	\\
Sens.		& $1.5\times 10^{-10}$	& $1.89\times 10^{-10}$	\\
Cand.		& 	1		& 	1		\\
\hline
${\cal B}$	&\multicolumn{2}{c|}{\PNNRESULT}			\\
\hline
\end{tabular}
\label{tab:pnn1}
\end{center}
\end{table}

 The search in the lower momentum region (Figure~\ref{fig:top7}), dubbed pnn2,
suffers from a larger background rate but has the advantage over pnn1 of greater
phase space and less loss due to $\pi^+N$ interactions. The pnn2
region also probes more of the  $ {\rm K}^+\to\pi^+\nu\bar\nu$ form factor.

\begin{figure}[htb]
\centering
\includegraphics*[width=65mm]{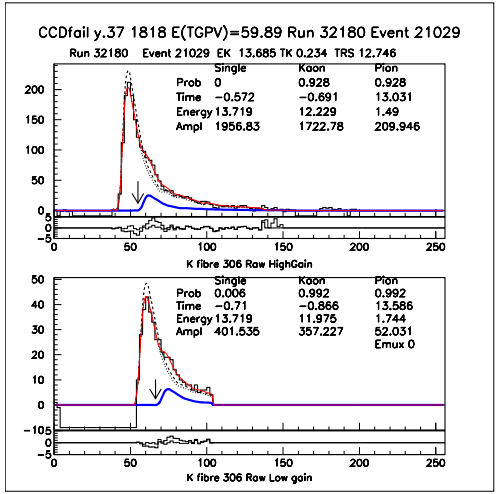}
\caption{Suppression of ${\rm K}^+\to\pi^+\pi^0$ scattering background.
The high-gain (upper) and low-gain (lower) CCD response is represented
by the histogram. The solid red (black dashed) lines represent
the results of a double-pulse (single-pulse) fit hypothesis. The solid
blue line represents the resolved second pulse of the
outgoing $\pi^+$. The vertical arrow indicates the expected time for
the second pulse based on the $\pi^+$ observed in the RS.
The thin histograms underneath the large plots show the 
residual distributions for the single- and double-pulse fit hypotheses.}
\label{fig:ccd}
\end{figure}

 The main pnn2 background is due to $K_{\pi2}$ decays where the $\pi^+$
is emitted along the beam axis and scatters into the RS. The scatter
destroys the back-to-back correlation of the  $\pi^+$ and  $\pi^0$
that allows the pnn1 analysis to suppress $K_{\pi2}$ background using the barrel
photon veto. For the $K_{\pi2}$-scatter background, at least one of
the  $\pi^0$ photons is directed into the beam region which is
necessarily less instrumented. Since the fibers in the target run along the
beam direction, the outgoing  $\pi^+$ can remain in the same fiber as the 
stopping  $ {\rm K}^+$ and its energy deposit signature can be obscured
by the larger  $ {\rm K}^+$ energy deposit. Some rejection of this background
is possible by using CCDs to digitize the energy deposit in each fiber every
2 ns. An example of the identification of the outgoing $\pi^+$ energy deposit
in a fiber traversed by the  $ {\rm K}^+$ is shown in Figure~\ref{fig:ccd}.

The preliminary pnn2 results are given in Table~\ref{tab:pnn2} with
a comparison to pnn1. The pnn2 search also suffers from 
background due to $K_{e4}$ (${\rm K}^+\to\pi^+\pi^- e^+\nu$) if
both the $\pi^-$ and $e^+$ are undetected and radiative  $K_{\pi2}$ 
decays that push the  $\pi^+$ into the pnn2 region. In contrast
to pnn1, ${\rm K}^+\to\mu X$ backgrounds are very small for pnn2 allowing
more acceptance by relaxing the criteria for  $\pi^+\to\mu^+\to e^+$
identification. The background to signal sensitivity of pnn2 in Table~\ref{tab:pnn2}
is approximately 20 times worse than that of pnn1.

\begin{table}[hbt]
\begin{center}
\caption{E787 results for the pnn1 and pnn2 search regions.
N(K) is the number of stopped $ {\rm K}^+$,
Acc. is the acceptance, Cand. is the number of observed signal candidates
and $\Br$ is $\Br({\rm K}^+\to\pi^+\nu\bar\nu)$.
The preliminary pnn2 limit is at 90\% C.L.}
\begin{tabular}{|lcc|}
\hline
\multicolumn{3}{|c|}{Backgrounds}	\\
Source 		& pnn2			& pnn1	\\
\hline
$K_{\pi 2}$	& {$1.029\pm0.227$}	& {$0.042{}^{+0.010}_{-0.011}$}	\\
Beam		& $0.066\pm0.047$	& $0.024\pm0.020$		\\
$K_{e4}$	& $0.052\pm0.041$	& NA	\\
$K_{\pi 2\gamma}$&$0.033\pm0.004$	& NA	\\
CEX		& $0.024\pm0.017$	& $0.026\pm0.011$		\\
$K\to\mu X$	& $0.016\pm0.011$	& {$0.054{}^{+0.044}_{-0.026}$}		\\
\hline
Total		& $1.22\pm0.24$		& $0.15\pm0.05$		\\
\hline
\hline		
N(K)		& $1.7\times10^{12}$	& $5.9\times10^{12}$\\
Acc.		& $0.835\times10^{-3}$	& $2.04\times10^{-3}$	\\	
Cand.		& 	1		& 	2	\\
\Br\		&  $ <22\times 10^{-10}$& \PNNRESULT\           \\
\hline
\end{tabular}
\label{tab:pnn2}
\end{center}
\end{table}

\section{E949}

 E949~\cite{ref:e949.expt} is an upgraded E787 detector designed to have increased
sensitivity to  $ {\rm K}^+\to\pi^+\nu\bar\nu$  in both pnn1 and pnn2 regions.
The upgrades improved photon veto hermeticity, both in the barrel and
beam regions, tracking resolution and DAQ for an 
increased duty factor. E949 accumulated $1.9\times 10^{12}$ stopped  ${\rm K}^+$
in an eleven-week run in 2002 and expects to have results for the pnn1
region by the end of 2003 with a sensitivity slightly less than E787 and about 20\%
of the E949 goal. Current E949 studies show that the upgraded detector would be
capable of achieving the E949 sensitivity goal of $<10^{-11}$. 
Unfortunately E949 is languishing due to a lack of funding
since the DOE terminated high energy physics running at the AGS in 2002.

\section{CKM}

 The CKM experiment~\cite{ref:ckm.expt} proposed at Fermilab has a sensitivity goal of $10^{12}$
corresponding to $\sim 100$  $ {\rm K}^+\to\pi^+\nu\bar\nu$ events at the SM
rate with a signal-to-background of $\sim 10$. CKM departs form the E787/E949 strategy
by seeking to measure  ${\rm K}^+$ decay in flight in a 
22 GeV$/c$, 50 MHz separated  ${\rm K}^+$ beam ($\sim 69\%$  ${\rm K}^+$ purity).
CKM plans to kinematically suppress backgrounds by $10^5$ with independent
measurements of the ${\rm K}^+$ and $\pi^+$ momentum and velocity vectors
with magnetic spectrometers and ring-imaging Cherenkov detectors. Additional
rejection of $10^7$ is achieved via a hermetic photon veto. The primary signal
region for CKM is similar to the pnn1 region of E787/E949. The pnn2 region may be more
accessible to CKM since the  $K_{\pi2}$-scatter background should not exist for
decay in flight.

\section{KOPIO}

 The situation with the decay $ {\rm K}^0_{\rm L}\to\pi^0\nu\bar\nu$ can be
summarized with two fourteen-year-old quotes.
``The process  $ {\rm K}^0_{\rm L}\to\pi^0\nu\bar\nu$  offers perhaps the
clearest window yet proposed into the origin of CP violation~\cite{ref:ll}.''
``Experimentally, the problems are perhaps best represented by the
statement that nobody has yet shown that a measurement of this
decay is absolutely impossible~\cite{ref:gilman}.''

\begin{figure}[htb]
\centering
\includegraphics*[width=65mm]{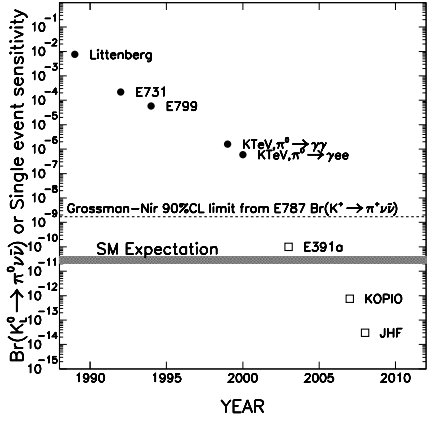}
\caption{Progress in the search for  $ {\rm K}^0_{\rm L}\to\pi^0\nu\bar\nu$.
The Grossman-Nir limit refers to \protect\cite{ref:grossman-nir}.}
\label{fig:k0.history}
\end{figure}

 Figure~\ref{fig:k0.history} shows the progress in the search for
 $ {\rm K}^0_{\rm L}\to\pi^0\nu\bar\nu$. The KTeV results with 
$\pi^0\to\gamma\gamma$~\cite{ref:ktev} utilized a well-collimated ``pencil''
neutral beam to constrain the  $ {\rm K}^0_{\rm L}$ decay point and 
measured the transverse momentum of the photon pair after vetoing
on all other particles. A similar approach is being followed
by KEK E391a. 

\begin{figure}[htb]
\centering
\includegraphics*[width=65mm]{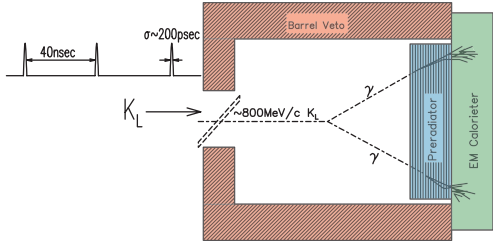}
\caption{A conceptual drawing of the KOPIO technique.}
\label{fig:kopio.cartoon}
\end{figure}

The KOPIO~\cite{ref:kopio.expt} experiment at BNL proposes to use a different
technique summarized pictorially in Figure~\ref{fig:kopio.cartoon}. The neutral
beam is produced in $\sim 250$ ps wide bunches every 40 ns and collimated
toward the KOPIO decay region. The time, direction and energy of the
two photons are measured
with a fine-grained preradiator (2$X_0$) and calorimeter (15$X_0$).
The $\pi^0$ is reconstructed from the momenta of the two photon
candidates with the constraint of a vertically narrow neutral beam
($100\times 5\ {\rm mrad}^2$). Applying the further constraint of the
 $\pi^0$ mass improves the measurement of the  $ {\rm K}^0_{\rm L}$
decay position and time and thus the  $ {\rm K}^0_{\rm L}$ velocity
thanks to the bunched beam. KOPIO will be able to kinematically suppress
backgrounds by working in the  $ {\rm K}^0_{\rm L}$ center-of-momentum system (CMS).
Additional background rejection from hermetic photon and charged particle
vetoes complement the kinematic rejection.

 Table~\ref{tab:kopio.bkgds} lists the backgrounds 
from \KL\ decays and the tools that will be used
to suppress them. Following the successful E787/E949 strategy, KOPIO is designed
to suppress each background with at least two independent criteria which 
will allow estimation of background rates from the data.

\begin{figure}[htb]
\centering
\includegraphics*[width=65mm]{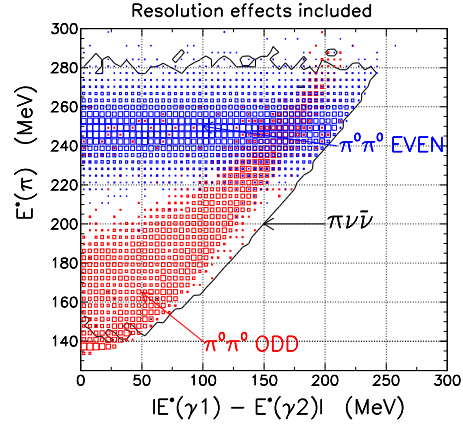}
\caption{The distribution of events from  $\KL\to\PZ\PZ$ decays in
the $E^*_\pi$ vs \DELTAE\ plane. The boxes represent the $\KL\to\PZ\PZ$
events and the solid envelope shows the range for   
${\rm K}^0_{\rm L}\to\pi^0\nu\bar\nu$ decays. Potential signal regions
are the sparsely populated regions near  the kinematic limits for
${\rm K}^0_{\rm L}\to\pi^0\nu\bar\nu$ and
the region near $\DELTAE=0$ between the $K_{\pi2}$-even and $K_{\pi2}$-odd
backgrounds.  $K_{\pi2}$-even(odd) denotes \PZ\ candidates
where the two photons are from the same(different) \PZ.}
\label{fig:kopio.kp2}
\end{figure}

 The main background is the CP-violating decay $\KL\to\PZ\PZ$ decay when two
photons escape the veto. Figure~\ref{fig:kopio.kp2} illustrates the power of
KOPIO's kinematic suppression of background. The expected background and signal
rates for the entire projected KOPIO three year running period are shown in 
Table~\ref{tab:kopio.results}. If the SM prediction is correct, KOPIO will observe
about 40  $ {\rm K}^0_{\rm L}\to\pi^0\nu\bar\nu$ events upon a background of 20 events
yielding a 20\% measurement of $\Br({\rm K}^0_{\rm L}\to\pi^0\nu\bar\nu)$ or a 
determination of the height of the UT to 10\%.

\begin{table}[hbt]
\begin{center}
\caption{ $ {\rm K}^0_{\rm L}$ background suppression in KOPIO.
The $ {\rm K}^0_{\rm L}$ branching fractions are given with respect
to the SM prediction for $\Br({\rm K}^0_{\rm L}\to\pi^0\nu\bar\nu)$.
PV (CV) is the photon (charged) veto.
 Even $\equiv$ both $\gamma$ from same \PZ,
 odd  $\equiv$ $\gamma$ from different \PZ\ ,
 \CHISQ $\equiv$ \CHISQ of fit of $\gamma$ 3-momenta to a common vertex,
 \MGG $ \equiv $ 2 photon invariant mass,
 $E^*_i \equiv $ energy in \KL\ rest frame, $i=\PZ,\gamma_1,\gamma_2$ and
 $\EMISS \equiv E(\KL) - E(\gamma_1) - E(\gamma_2)$.}
\begin{tabular}{|lclcc|}
\hline 
\KL\ Decay   & $\BR/3\times10^{-11}$ & Kinematic &  PV   &  CV \\
\hline 
\hline 
\PZ\PZ\ even   &  $3.1\times10^7$    & \ESTAR                        &  $\surd \surd$&             \\
\PZ\PZ\ odd    & $3.1\times10^7$     &        \DELTAE                  & $\surd \surd$ &             \\
               &                     &                  \MGG\          &               &             \\
\hline 
\KEPG\    & $1.2\times10^8$     & \MGG,\ \CHISQ& $\surd $     &    $\surd $            \\
\hline  
\KPPP\    & $4.2\times10^9$    &\ESTAR,\ \EMISS           &             &  $\surd \surd$            \\
\hline\hline  
\KEPP\    & $1.7\times10^6$  & \ESTAR & & $\surd\surd$ \\
\hline 
\KZZZ\ & $7.0\times10^9$ &\ESTAR & $\surd\surd\surd$ & \\
\hline 
\KPGG\ & $5.6\times10^4$ &       & $\surd\surd$& \\
\hline 
\KGG\  & $2.7\times10^7$ & \MGG,\ \ESTAR & & \\
\hline 
\end{tabular} 
\label{tab:kopio.bkgds}
\end{center} 
\end{table} 

\begin{table}[hbt]
\begin{center}
\caption{ The expected signal and background rates in KOPIO
assuming the SM prediction for $\Br({\rm K}^0_{\rm L}\to\pi^0\nu\bar\nu)$.}
\begin{tabular}{lc}
Process				& Events	\\
\hline
$ {\rm K}^0_{\rm L}\to\pi^0\nu\bar\nu$ 	at SM rate		& 40	\\
\hline
$\KL\to\PZ\PZ$			& 12.4	\\
$\KL\to\KEPG$ 			& 4.5	\\
$\KL\to\pi^-\pi^+\pi^0$		& 1.7	\\
$\KL\to\pi^\pm e^\mp\nu$	& 0.02	\\
$\KL\to\gamma\gamma$		& 0.02	\\
$\Lambda\to\pi^0n$		& 0.01	\\
Interactions ($nN\to\PZ X$)	& 0.2	\\
Accidentals			& 0.6	\\
\hline\hline
Total Background		& 19.5	\\
\end{tabular}
\label{tab:kopio.results}
\end{center} 
\end{table}

\section{Summary}

\begin{figure}[htb]
\centering
\includegraphics*[width=65mm]{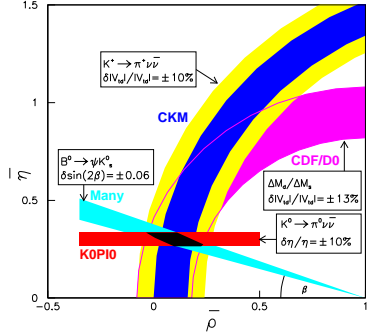}
\caption{The expected precision on the apex of the unitarity triangle
from measurements of $\sin2\beta$, $\Delta m_s$ and the $ {\rm K}\to\pi\nu\bar\nu$
branching fractions assuming all results are consistent with current
SM predictions.}
\label{fig:future.sm}
\end{figure}

\begin{figure}[htb]
\centering
\includegraphics*[width=65mm]{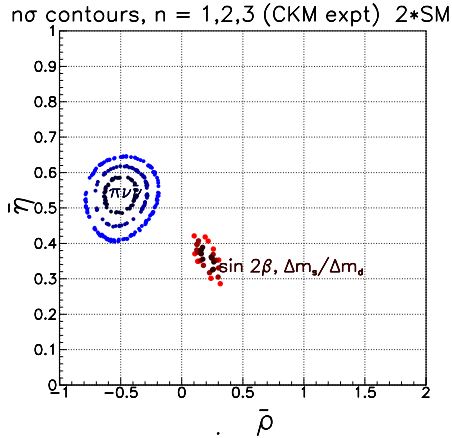}
\caption{The expected precision on the apex of the unitarity triangle
from measurements of $\sin2\beta$, $\Delta m_s$ and the $ {\rm K}\to\pi\nu\bar\nu$
branching fractions assuming the former two measurements  are consistent with current
SM predictions whilst the latter two are at twice the SM predictions.
The contours represent 1, 2 and 3 standard deviations.}
\label{fig:future.2sm}
\end{figure}

 The measurement of the branching fraction of both  $ {\rm K}\to\pi\nu\bar\nu$
decay modes with the design sensitivity of CKM and KOPIO could yield 
confidence level contours
for the apex of the UT similar to those 
shown in Figures~\ref{fig:future.sm}~\cite{ref:ckm.expt} and \ref{fig:future.2sm}~\cite{ref:sm.pred}.
Clearly such measurements would be able to test the precise predictions
of the SM for the fundamental parameters of the CKM matrix.

\cleardoublepage 

\subsection{Acknowledgements}
 I would like to thank Gino Isidori for inviting me to the workshop and
 Nicol\`o Cartiglia for a careful reading of this paper.
Peter Cooper, George Redlinger, Steve Kettell, Laurence Littenberg, Ken Nelson,
Hogan Nguyen and Bob Tschirhart helped to prepare my presentation by 
graciously providing figures and comments. I also wish to thank the
E949, CKM and KOPIO collaborations.

\end{document}